\documentclass{iau} 
\usepackage{graphicx}
\usepackage{subfigure}
\usepackage{amsmath}
\usepackage{amssymb}
\usepackage{bm}
\usepackage{url}

\title[] 
{Exploiting simultaneous multi-frequency observations to probe polar-cap processes}

\author[Yogesh Maan]   
{Yogesh Maan}
\affiliation{ASTRON, Netherlands Institute for Radio Astronomy\\
Postbus 2, 7990 AA Dwingeloo, the Netherlands \\ email: {\tt maan@astron.nl}}

\pubyear{2017}
\volume{337}  
\jname{Pulsar Astrophysics -­ The Next 50 Years}
\editors{P. Weltevrede, B.B.P. Perera, L. Levin Preston \& S. Sanidas, eds.}

\begin{document}
\maketitle
\begin{abstract}
Sub-pulse drifting has been regarded as one of the most insightful aspects
of the pulsar radio emission. The phenomenon is generally explained with a
system of emission sub-beams rotating around the magnetic axis, originating from
a carousel of sparks near the pulsar surface (the carousel model). Since the
observed radio emission at different frequencies is generated at different
altitudes in the pulsar magnetosphere, corresponding sampling of the carousel
on the polar cap differs slightly in magnetic latitude. When this aspect is
considered, it is shown here that the carousel model predicts important
observable effects in multi-frequency or wide-band observations. Also presented
here are brief mentions of how this aspect can be exploited to probe the
electrodynamics in the polar cap by estimating various physical quantities,
and correctly interpret various carousel related phenomena, in addition to
test the carousel model itself.
\end{abstract}

\firstsection 
\section{Carousel model and radius-to-frequency mapping of radio emission}
A considerable fraction of pulsars exhibit a systematic modulation of
intensity within individual pulses. The modulation appearing in the form
of \textit{components within single pulses, i.e., sub-pulses}, generally
drifts in pulsar's rotation phase from pulse-to-pulse. This fascinating
phenomenon, called sub-pulse drifting, is generally attributed to the
presence of a system of emission sub-beams stemmed from a carousel of
`sparks' near the pulsar surface rotating around the magnetic axis (the
carousel model; \cite[Ruderman \& Sutherland 1975]{RS75}).
\par
Radius-to-frequency mapping (RFM; \cite[Cordes 1978]{Cordes78}) suggests the
observed radio emission at different frequencies to be generated at different
altitudes in the pulsar magnetosphere. Furthermore, foot-points of the field
lines that give
rise to the emission at different altitudes have slightly different magnetic
colatitudes at the polar cap. Thus, different frequencies sample slightly
offset slices of the polar cap, and hence, the carousel at its
different rotation phases. This aspect implies important observable signatures
of the carousel in multi-frequency (e.g., \cite[Maan et al. 2013]{Maan13})
and wide-band observations, as briefed below.
\section{Multi-frequency observational consequences}
To demonstrate the observational effect of the above aspect at different
frequencies, two pulse-sequences were generated (at presumed
frequencies of 240 and 610\,MHz) from a simulated carousel of 6 sparks and
using an empirical relationship for RFM. For a sightline
non-tangential to the emission beam, such a carousel would produce two
pulse-components. The different sampling of the carousel at its slightly
different circulation phases reflects as an offset in sub-pulse modulation
phase under a given pulse-component at different frequencies. Moreover, such a
phase offset will be equal in magnitude but opposite in direction for the
two components (see Figure 1). The magnitude of the phase offset is primarily
contributed by the difference in magnetic azimuths corresponding to the
pulse-longitudes of the given pulse-component at different frequencies.
A relatively smaller, second order contribution comes from the circulation of the
carousel during the time corresponding to the difference in
pulse-longitudes of the same component at different frequencies. 
\par
For a known viewing geometry and measurable sub-pulse modulation period, P$_3$,
the above offset in sub-pulse phase is a completely deterministic quantity and can be
exploited to determine the carousel circulation period, potentially resolve
the P$_3$-aliasing issue even for non-tangential sightlines, and test the
carousel model itself using simultaneous dual/multi-frequency observations.
The above phase offset, when translated to the corresponding delay, can be further
exploited to directly measure the angular velocity of the carousel using the
pulse intensity fluctuations observed \emph{simultaneously} at two or more
radio frequencies, even if no sub-pulse
modulation is apparent (due to, e.g., an irregular or unstable carousel).
This phase-offset also has important implications when interpreting
multi-frequency occurrences of pseudo-nulls, sub-pulse modulations in
orthogonal polarization modes, exploring any twist in the field line
geometry using sub-pulse modulations (\cite[Maan \& Deshpande 2014]{MD14}), etc.
\par
Summarizing, the carousel model along with the RFM predicts offsets
in sub-pulse modulation phases under a given pulse-component when observed
simultaneously at different frequencies for viewing geometries non-tangential
to the emission beam.
\begin{figure*}
\centering
\includegraphics[width=0.71\textwidth,angle=0]{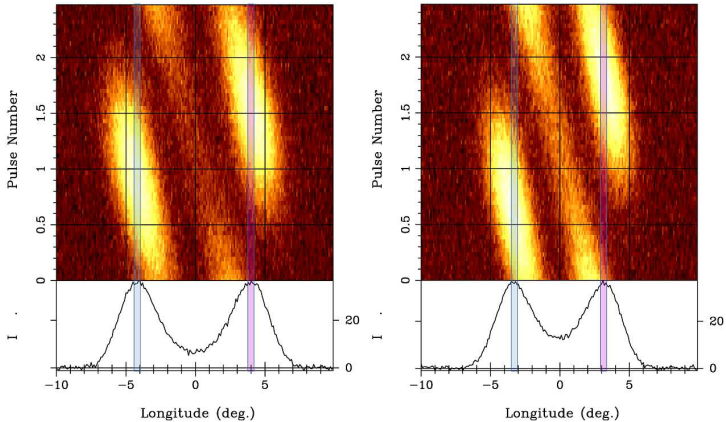}
\caption{Example of two pulse-sequences simulated at 240 and 610\,MHz and
folded over the sub-pulse modulation period. The offset in sub-pulse modulation
phase under the corresponding components (compare, for example, the light blue
shaded regions) is clearly evident.}
\label{modfolds}
\end{figure*}
\acknowledgements
The research leading to these results has received funding from the
European Research Council under the European Union's Seventh Framework
Programme (FP/2007-2013) / ERC Grant Agreement n. 617199.

\end{document}